\documentclass[11pt]{article}
\usepackage{graphicx}
\usepackage[latin1]{inputenc}
\usepackage{epsfig}
\usepackage{amsmath,amsfonts,latexsym, amssymb}
\usepackage{array}
\usepackage{psfrag}
\newcommand{\card}{\text{card}}

\newtheorem{satz}{Theorem}[section]
\newtheorem{defi}[satz]{Definition}

\newtheorem{ob}[satz]{Observation}

\newtheorem{res}[satz]{R\'esum\'e}
\newtheorem{conj}[satz]{Conjecture}

\newcommand{\tit}{\textit}

\newcommand{\Z}{\mathbb{Z}}

\begin{document}
\thispagestyle{empty}
\begin{center}
\vspace*{0.5cm}

{\LARGE{\bf Emergent Properties in Structurally Dynamic Disordered
    Cellular Networks}} 

\vskip 1.0cm

\parbox[t]{0.48\textwidth}{
{\large {\bf Thomas Nowotny }} 

\vskip 0.5 cm 

Institute for Nonlinear Science \\ 
University of California San Diego \\ 
Mail code 0402 \\
9500 Gilman Dr. \\ 
La Jolla, CA 92093-0402 \quad USA \\
(E-mail: tnowotny@ucsd.edu)
}
\hfill
\parbox[t]{0.48\textwidth}{
{\large {\bf Manfred Requardt }} 

\vskip 0.5 cm 

Institut f\"ur Theoretische Physik \\ 
Universit\"at G\"ottingen \\ 
Friedrich-Hund-Platz 1 \\ 
37077 G\"ottingen \quad Germany\\
(E-mail: requardt@theorie.physik.uni-goettingen.de)
}

\end{center}

\vspace{1 cm}
\begin{abstract}
We relate structurally dynamic cellular networks, a class of models we
developed in fundamental space-time physics, to SDCA, introduced some
time ago by Ilachinski and Halpern. We emphasize the crucial property
of a non-linear interaction of network geometry with the matter degrees
of freedom in order to emulate the supposedly highly erratic and
strongly fluctuating space-time structure on the Planck scale. We
then embark on a detailed numerical analysis of various large scale
characteristics of several classes of models in order to understand
what will happen if some sort of macroscopic or continuum limit is
performed. Of particular relevance in this context is a notion of
network dimension and its behavior in this limit. Furthermore, the
possibility of phase transitions is discussed.
\end{abstract} 

\newpage
\setcounter{page}{1}
\section{Introduction}
In the beautiful book \cite{Ila1} the title of chapter 12 reads: ''Is
Nature, underneath it All, a CA?''. Such ideas have in fact been
around for quite some time (cf.\ e.g.\
\cite{Zuse},\cite{Feynman},\cite{Fredkin} or \cite{Finkelstein}, to
mention a few references). A little bit later 't Hooft analysed the
possibility of deterministic CA underlying models of quantum field
theory or quantum gravity (\cite{Hooft1} and \cite{Hooft2} are two
examples from a long list of papers). For more detailed historical
information see \cite{Ila1} or \cite{Wolfram}. A nice collection of
references can also be found in \cite{Svozil}. However, we would like
to issue a warning against an overly optimistic attitude. While we
share the general philosophy uttered in these works, there are some
subtle points as 't Hooft remarks correctly(\cite{Hooft3}). It is no
easy task to incorporate something as complex as the typical
\tit{entanglement structure} of quantum theory into the, at first
glance, quite simple and local CA-models. We would like to emphasize
that it is not sufficient to somehow simulate or reproduce these
quantum phenomena numerically on a computer or CA. What is actually
called for is a \tit{structural} isomorphism between those phenomena
and corresponding \tit{emergent} phenomena on CA. This problem has
been one of the reasons underlying our interest in CA having a
fluctuating time-dependent geometry (see below). We recently observed
that ideas about the discrete fine structure of space-time similar to
our own working philosophy have been uttered in chapt. 9 of
\cite{Wolfram}, in particular concerning the existence of what we like
to call \tit{shortcuts} or \tit{whormhole structure}.

Still another interesting point is discussed by Svozil
(\cite{Svozil2}), i.e. the well-known problem of \tit{species
  doubling} of fermionic degrees on regular lattices, which, as he
argues, carries over to CA. Among the various possibilities to resolve
this problem he suggests a kind of \tit{dimensional reduction}
(``dimensional shadowing''), which leads in the CA one is actually
interested in, to \tit{non-local} behavior (see also \cite{Ila1}
p.649ff). It is perhaps remarkable that, motivated by completely
different ideas, we came to a similar conclusion concerning the
importance of non-local behavior (cf. \cite{Requ4}, see also
\cite{Requloch}).

While presently the discussion in the physics community, when it comes
to the high-energy end of fundamental physics, is dominated by string
theory and/or loop quantum gravity, frameworks which are in a
conceptual sense certainly more conservative, we nevertheless regard
an approach to these primordial questions via \tit{networks} and/or CA
as quite promising. In contrast to the above-mentioned (more
conservative) approaches which start from continuum physics and hope
to detect discrete space-time behavior at the end of the analysis (for
example by imposing quantum theory as a quasi God-given absolute
framework on the underlying structures over the full range of scales),
we prefer a more \tit{bottom-up-approach}. One of our reasons for this
preference is that we do \tit{not} believe that quantum theory holds
sway unaltered over the many scales addressed by modern physics down
to the pristine Planckian regime. Like 't Hooft, we regard quantum
theory rather as a kind of effective intermediate framework, which
emerges from some more primordial structure of potentially very
different nature. We start from some underlying dynamic, discrete and
highly erratic network substratum consisting of (on a given scale)
irreducible agents interacting (or interchanging pieces of
information) via elementary channels. On a more macroscopic (or,
rather, mesoscopic) scale, we then try to reconstruct the known
continuum structures as emergent phenomena via a sequence of
\tit{coarse graining} and/or \tit{renormalisation} steps (see
\cite{Requ3} and \cite{Requ2}).

While CA have been widely used in modeling complex behavior of molecular
agents and the like (a catchword being \tit{artificial life} or Conway's
\tit{game of life}; for a random selection see e.g. \cite{Ila1},
\cite{Kauffmann1}, \cite{Kauffmann2}, \cite{Kauffmann3}, \cite{Langton} or
\cite{Gardner1}), papers on the more pristine and remote regions of
Planck-scale physics are understandably less numerous.

When we embarked on such a programme in the early nineties of the last
century, we soon realized that the ordinary framework of CA, typically
living on fixed and quite regular geometric arrays, appears to be far to
rigid and regular in this particular context. In order to implement the
lessons of general relativity we have to make their structure dynamical, that
is, not only the local states on the vertices of the lattice but also the
local states attached to the links need to be dynamic. A fortiori, we would
like the whole wiring diagram of links to be ``clock-time dependent''. To put
it briefly: \tit{matter} shall act on \tit{geometry} and vice versa, where
we, tentatively, associate the pattern of local vertex states with the matter
distribution and the geometric structure of the network with \tit{geometry}.

Our first task therefore is to turn both the site and the link states
into fully dynamical degrees of freedom, which mutually depend on each other
in a dynamical and local way. Furthermore, all this is assumed to happen on
very irregular arrays of nodes and links which dynamically arrange themselves
according to some given evolution law. Then the hope is, that under certain
favorable conditions, the system will undergo a (series of) phase transition(s)
from, for example, a disordered chaotic initial state into a kind of
macroscopically ordered, extended pattern, which may be associated with a
classical continuum space-time with some matter living in it.

One of our first (published) papers, in which we implemented such a
programme was \cite{Requ1}, see also \cite{Nowrequ1}; for the notion
of dimension of such irregular structures see \cite{Nowrequ2}. We
carefully inspected the literature known to us on CA at the time of
writing those papers, but only several years later, when one of the
authors (M. Requardt) had the pleasure to review the book by
Ilachinski, we became aware of slightly earlier related ideas
developed by Ilachinski and Halpern (see e.g. \cite{Ila1} or
\cite{Halpern} for reviews and further references). In the following
sections we are going to relate these two originally independent
approaches to each other and discuss the behavior of two interesting
dynamical network models we employed and studied in greater
detail. Furthermore, we give an overview of an extensive numerical and
computational analysis of these model systems.
\section{A Comparison of SDCA and our Dynamic Cellular Networks}
\subsection{SDCA}
SDCA have been introduced by Ilachinski and Halpern and are
straightforward generalisations of CA (for a more recent application
see e.g. \cite{Sanz}). In the simplest cases they are placed (as most
of CA) on a finite or infinite regular grid, e.g. $\Z^d$. The
generalisation consists of the assumption that also the links,
connecting the sites of the lattice, can be created and deleted
according to a local law. More properly, we have link variables,
$l_{xy}$, attaining the value 1 if site $x$ is linked to site $y$ and
being zero otherwise.

In this context it is of course of great relevance which sites can be linked
at all. In \cite{Ila1} or \cite{Halpern}, for example, links of the original
background lattice $\Z^2$ belong to this pool together with diagonal links to
the next-nearest neighbors. In the respective examples some start
configuration is chosen on the Euclidean background lattice and one can
observe, in the course of clock time, the emergence of additional diagonal
links and the subsequent deletion of some of them, as well as deletion and
reinsertion of the original horizontal and vertical links . The local
dynamical rule guarantees that only links connecting nearest or next-nearest
neighbors participate in the process (cf. e.g. section 8.3 in
\cite{Ila1}). This entails that the change of the wiring diagram proceeds
still in a rather local and orderly way with respect to the initial Euclidean
lattice.

These restrictions are of course not necessary. In general, CA can be defined
on an undirected graph. At each site $x$ we have attached a site state $s_x$,
being capable of attaining some discrete values (typically $s_x\in \{0,1\}$)
while link states $l_{xy}$ can have the values $0$ or $1$. In SDCA the wiring
diagram, i.e., the distribution of links, is now also a dynamical evolving
structure. A local law, being independent of the constantly varying wiring
diagram can be formulated by employing the natural graph distance metric
given by
\begin{equation}d(x,y):=\min\{l_{\gamma}\} \end{equation} with
$\gamma$ a path, connecting the sites $x$ and $y$ and $l_{\gamma}$ its
length, i.e., the number of links along the path (this distance being
infinite if the sites are in disconnected pieces of the network). With this
metric, the graph becomes a discrete metric space. Ball neighborhoods around
a site $x$ are then defined by
\begin{equation}B_r(x):=\{y,d(x,y)\leq r\} \end{equation}
(cf. e.g. \cite{Nowrequ2} or \cite{Requ2}).

For convenience we introduce some notation. The underlying time
dependent lattice (the wiring diagram) is denoted by $L_t$.
$s_x,l_{xy}$ (or $s_i,l_{ik}$) designate the local site or link states
(in the simplest case $s_x\in \{1,0\}$, $l_{xy}\in \{1,0\}$). $N(x)$
is a certain neighborhood of sites and links about the site $x$. A
classical CA is given by a local dynamical law or rule, i.e., a map
from some $N(x)$ to $S$, the state space at site $x$. Typically the
type of neighborhood and the local rule are chosen to be the same over
the full lattice.

Things become a little bit more complicated if the wiring diagram is chosen
to also become (clock) time dependent. In that case it is more reasonable to
define the neighborhoods by the distance metric, i.e., choose some
$B_r(x)$. Note that now the actual site and link content of $B_r(x)$ is time
dependent, while the definition of the neighborhood can be given in a time
independent form. From a mathematical point of view we could formulate rather
arbitrary local rules, but physics has taught us to avoid too artificial or
cumbersome rules, which depend on rather {\em ad hoc} assumptions. So, quite
reasonable laws appear to be \tit{totalistic} or \tit{outer-totalistic}
rules, which act on the sum of site states and/or link states in the ball
$B_r(x)$ with a possible particular role played by $s_x$ itself.

In the general case, the dynamics is given by a pair of local laws:
\begin{align}s_x^{(t+1)} & =F_1(s^t_{x'}\in B_r(x),l^t_{x'y'}\in
  B_r(x))\\l^{(t+1)}_{xy} & =F_2(s^t_{x'}\in B_{r'}(l^t_{xy}),l^t_{x'y'}\in
  B_{r'}(l^t_{xy})),
\end{align}
in which we have been a bit sloppy in order not to overburden the formulas
with too many indices. To get an idea how this scheme works in concrete
examples, see sect. 8.8 of \cite{Ila1} or \cite{Halpern} or the sections
below. \\[0.3cm]

Remark: We would like to emphasize again, that, in the typical
examples given above, link deletion or creation is restricted to
nearest or next-nearest neighbors with respect to the background
lattice (e.g. $\Z^2$). The lattice evolution is hence still quite
regular. This is of some relevance in comparison to our cellular
networks, which are capable of developing both local and translocal
connections with respect to some reference space.
\subsection{Our Dynamic Cellular Networks}
Our networks are defined on general graphs, $G$, with $V(G)$ the set
of vertices (sites or nodes) and $E(G)$ the set of edges (links or
bonds). The local site states can assume values in a certain discrete
set. In the examples we have studied, we follow the philosophy that the
network should be allowed to find its typical range of states via the
imposed dynamics. That is, we allow the $s_i$ to vary in principle
over the set $q \cdot \Z$, with $q$ a certain discrete quantum of
information, energy or whatever. The link states can assume the values
$J_{ik}\in \{-1,0,+1\}$ (we are assuming here the notation $J_{ik}$
instead of $l_{ik}$ as we regard the links as representing a kind of
elementary coupling).

Viewed geometrically we associate the states $J_{ik}= +1,-1,0$ with directed
edges pointing from site $x_i$ to $x_k$, or the other way around, or, in the
last case, with a non-existing edge. That is, at each clock time step,
$t \cdot \tau$ ($\tau$ an elementary quantum of time), we have as underlying
substratum a time dependent \tit{directed} graph, $G_t$.  Our physical idea
is that at each clock time step an elementary quantum $q$ is transported
along each existing directed edge in the indicated direction.

To implement our general working philosophy of mutual interaction of overall
site states and network geometry, we now describe some particular network
laws, which we investigated in greater detail (see the following section). We
mainly considered two different classes of evolution laws for vertex and edge
states:
\begin{itemize}
\item Network type I
  \begin{align}
    s_i(t +1)& = s_i(t)+\sum_k J_{ki}(t)\\
    J_{ik}(t+1)& = \text{sign}(\Delta s_{ik})\;\text{for}\;|\Delta s_{ik}| \geq \lambda_2 \vee
    \big(|\Delta s_{ik}| \geq \lambda_1 \wedge J_{ik}(t) \neq 0 \big) \\
  J_{ik}(t+1)& =  0 \quad \text{o.w.}
     \end{align}
\item Network type II
  \begin{align}
    s_i(t +1)=& s_i(t)+\sum_k J_{ki}(t)\\
  J_{ik}(t+1)& =\text{sign}(\Delta s_{ik})\;\text{for}\; 0 < |\Delta s_{ik}| < \lambda_1 \vee
    \big(0 < |\Delta s_{ik}| < \lambda_2 \wedge J_{ik}(t) \neq 0 \big) \\
  J_{ik}(t+1)& = J_{ik}(t)\;\text{for}\;\Delta s_{ik} = 0 \\
  J_{ik}(t+1)& =  0 \quad \text{o.w.}
   \end{align}
\end{itemize}
where $\Delta s_{ik}= s_{i}(t) - s_k(t)$ and $\lambda_2 \geq
\lambda_1\geq 0$. We see that in the first case, vertices are
connected that have very different internal states, leading to large
local fluctuations, while for the second class, vertices with similar
internal states are connected.

We proceed by making some remarks in order to put our approach into the
appropriate context.\\[0.3cm]
Remarks:\hfill\\
\begin{enumerate}
\item It is important that, generically, laws, as introduced above, do not
  lead to a reversible time evolution, i.e., there will typically be
  \tit{attractors} or \tit{state-cycles} in total phase space (the overall
  configuration space of the node and bond states). On the other hand, there
  exist strategies (in the context of cellular automata!) to design
  particular \tit{reversible} network laws (cf. e.g. \cite{Toffoli}) which
  are, however, typically of second order. Usually the existence of
  attractors is considered to be important for \tit{pattern formation}. On
  the other hand, it may suffice that the phase space, occupied by the
  system, shrinks in the course of evolution, that is, that one has a flow
  into smaller subvolumes of phase space.
\item In the above class of laws a direct bond-bond interaction is not
  yet implemented. We are prepared to incorporate such a (possibly
  important) contribution in a next step if it turns out to be
  necessary. In any case there are not so many ways to do this in a
  sensible way. Stated differently, the class of possible, physically
  sensible interactions, is perhaps not so large.
\item We would like to emphasize that the (undynamical) clock-time,
  $t$, should not be confused with the ``true'' \tit{physical time},
  i.e., the time operationally employed on much coarser scales. The
  latter is rather supposed to be a collective variable and is
  expected (or hoped!) to emerge via a cooperative effect. Clock-time
  may be an \tit{ideal element}, i.e., a notion which comes from
  outside, so to speak, but -- at least for the time being -- we have
  to introduce some mechanism, which allows us to label consecutive
  events or describe \tit{change}.
\end{enumerate}
The following observation we make because it is relevant if one follows the
general spirit of modern high energy physics.
\begin{ob}[Gauge Invariance] The above dynamical law depends nowhere on the 
absolute values of the node charges but only on their relative
differences. By the same token, charge is nowhere created or destroyed. We have
\begin{equation}\Delta(\sum_{i \in I} s_i)=0\end{equation}
where, for simplicity, we represent the set of sites by their set of
indices, $I$, and $\Delta$ denotes the difference between consecutive
clock-time steps. Put differently, we have conservation of the global
node charge. To avoid artificial ambiguities we can, e.g., choose a
fixed reference level and take as initial condition the constraint
\begin{equation}\sum_{i \in I} s_i= 0\end{equation} 
\end{ob}

We conclude this subsection by summarizing the main steps of our working
philosophy.
\begin{res}
  Irrespective of the technical details of the dynamical evolution law
  under discussion, the following, in our view crucial,
  principles should be emulated in order to match fundamental requirements
  concerning the capability of {\em emergent} and {\em complex} behavior.
\begin{enumerate}
\item As is the case with, say, gauge theory or general relativity,
  our evolution law on the surmised primordial level should implement
  the mutual interaction of two fundamental substructures, put
  sloppily: ``{\em geometry}'' acting on ``{\em matter}'' and vice
  versa, where in our context ``{\em geometry}'' is assumed to
  correspond in a loose sense with the local and/or global bond states
  and ``{\em matter}'' with the structure of the node states.
\item By the same token, the alluded {\em self-referential} dynamical
  circuitry of mutual interactions is expected to favor a kind of
  {\em undulating behavior} or {\em self-excitation} above a return
  to some uninteresting `{\em equilibrium state}' as is frequently
  the case in systems consisting of a single component which directly
  feeds back on itself. This propensity for the `{\em autonomous}'
  generation of undulation patterns is in our view an essential
  prerequisite for some form of ``{\em protoquantum behavior}'' we
  hope to recover on some coarse grained and less primordial level of
  the network dynamics.
\item In the same sense we expect the overall pattern of switched-on and
 -off bonds to generate a kind of ``{\em protogravity}''.
\end{enumerate}
\end{res}
\section{Numerical Studies}
We now put our two cellular network models on a simplex graph with $n$
vertices $x_i$ and edges $e_{ij}$, $i,j \in \{1, \ldots, n\}$. More
specifically, the maximally possible number of edges is $n(n-1)/2$. We
choose such a simplex graph as initial geometry. As an initial
distribution for vertex states (seed) we choose a uniform (random)
distribution scattered over the interval $\{-k, -k+1, \ldots k-1,
k\}$. In an early state of the work we also used other (initial)
distributions as well but we did not find any significantly different
results. The initial values for edge states $J_{ik}$ were chosen from
$\{-1,1\}$ with equal probability $1/2$. In other words, our initial
state is a maximally entangled nucleus of vertices and edges and the
idea is to follow its unfolding under the imposed evolution laws. In a
sense, this is a scenario which tries to imitate the \tit{big bang}
scenario. The hope is, that from this nucleus some large-scale patterns
may ultimately emerge for large clock-time.

\begin{figure}
  \psfrag{ylabel1}{\small $\langle \Delta s_{ik} \rangle_{ik}$}
  \psfrag{ylabel2}{\small $\quad \sigma\big(\Delta s_{ik}\big)$}
  \psfrag{ylabel3}{\small $\card \{J_{ik} = 0\}$}
  \includegraphics[width= \textwidth]{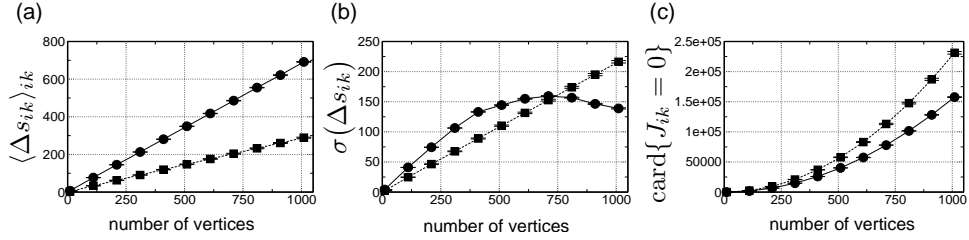}
  \caption{Examples of the observed dependence of averaged quantities on
  network size. Most quantities scale linearly with network size for both
  types of network laws like illustrated in (a). For some quantities the
  scaling is unclear, as for $\sigma(\Delta s_{ik})$ for network type I shown
  in (b). The number of ``off'' bonds shown in (c) scales with the total
  number of bonds, i.e. with $n^2$. The circles correspond to networks of
  type I, the squares to networks of type II. \label{lineardep}}
\end{figure}

For numerical investigations, the size of the CA is by necessity rather
limited. To obtain an estimate for properties of the large networks, which we
are ultimately interested in, we simulated networks of increasing size and
tried to extrapolate the expected properties for larger networks. The average
of all vertex states is approximately $0$ by construction and the the sum of
all temporal changes of vertex states is exactly $0$. For most other
properties we found that the average over the width of the initial vertex
state distribution, over $\lambda_1$ and $\lambda_2$, and over specific
realizations of initial conditions as well as time, has a linear dependence
of the given property on the network size $n$. Figure \ref{lineardep}(a)
shows an example and table \ref{propertytable} summarizes the observed
dependencies. A few of the quantities did not show linear dependencies, see
figure \ref{lineardep}(b) and (c). While the standard deviation of spatial
fluctuations has an unknown dependency on network size (\ref{lineardep}(b)),
the number of ``off'' bonds clearly scales with the number of total edges,
i.e., with $n^2$. This suggests that a random graph approximation with
constant probability for ``off'' edges might be possible.

\begin{table}
  \footnotesize
  \begin{tabular}{l|l|ccccc}
    \multicolumn{1}{c|}{\raisebox{-0.5em}{property}} &
    \multicolumn{1}{c|}{\raisebox{-0.5em}{law}} &
    \multicolumn{5}{c}{scaling of} \\[-0.5em] 
    & & card & min & mean & sigma & max \\
    \hline
    \raisebox{-0.5em}{vertex states $s_i(t)$} & law I 
    & $0.3 \, n$ & $-0.85 \, n$ & small & $0.85 \, n$ & $0.5 \, n$ \\[-0.5em]
     & law II
    & $0.3 \, n$ & $-n$ & small & $0.58 \, n$ & $n$  \\
    \hline
    \raisebox{-0.5em}{temporal fluctuations $\Delta s_i(t)$} & law I
    & $0.15 \, n$ & $-0.78 \, n$ & $0$ & $0.53 \, n$ & $0.78 \, n$ \\[-0.5em]
     & law II
    & $0.54 \, n$ & $-0.47 \, n$ & $0$ & $0.26 \, n$ & $0.47 \, n$ \\
    \hline
    \raisebox{-0.5em}{spatial fluctuations $\Delta s_{ik}(t)$} & law I
    & $1.4 \, n$ & $0$ & $0.55 \, n$ & ? & $1.7 \, n$ \\[-0.5em]
     & law II
    & $1.05 \, n$ & $0$ & $0.29 \, n$ & $0.21 \, n$ & $1.10 \, n$ \\
    \hline
    \raisebox{-0.5em}{spatial fluctuations $\Delta s_{ik}(t)$, $J_{ik} \neq 0$} & law I 
    & $1.2 \, n$ & $0.1 \, n$ & $0.7 \, n$ & ? & $1.7 \, n$ \\[-0.5em]
     & law II
    & $1.05 \, n$ & $0$ & $0.29 \, n$ & $0.21 \, n$ & $1.10 \, n$ \\
    \hline
    \raisebox{-0.5em}{vertex degrees $deg_i(t)$} & law I 
    & $0.04 \, n$ & $0$ & $0.7 \, n$ & $0.05 \, n$ & $n-1$ \\[-0.5em]
     & law II 
    & $0.23 \, n$ & $0.31 \, n$ & $0.55 \, n$ & $0.08 \, n$ & $0.68 \, n$ \\
    \hline
    \raisebox{-0.5em}{temporal fluctuations $\Delta deg_i(\tau)$} & law I
    & $0.03 \, n$ & $-0.12 \, n$ & small & $0.04 \, n$ & $0.12 \, n$ \\[-0.5em]
     & law II
    & $0.24 \, n$ & $-0.17 \, n$ & small & $0.07 \, n$ & $0.18 \, n$ \\
    \hline
    \raisebox{-0.5em}{spatial fluctuations $\Delta deg_{ik}(t)$} & law I
    & $0.1 \, n$ & $0$ & $0.05 \, n$ & ? & $0.22 \, n$ \\[-0.5em]
     & law II
    & $0.36 \, n$ & $0$ & $0.09 \, n$ & $0.08 \, n$ & $0.37 \, n$ \\
    \hline
    \raisebox{-0.5em}{spatial fluctuations $\Delta deg_{ik}(t)$, $J_{ik} \neq 0$} & law I
    & $0.08 \, n$ & small & $0.08 \, n$ & ? & $0.25 \, n$ \\[-0.5em]
     & law II
    & $0.32 \, n$ & $0$ & $0.07 \, n$ & $0.06 \, n$ & $0.34 \, n$
    \end{tabular}
  \normalsize
  \caption{Scaling laws for averaged properties of the two network
  types. Most quantities scale linearly with the number of nodes $n$. Some
  scale with the square of $n$, while others have an unknown scaling
  behaviour, denoted by a question mark in the table. \label{propertytable}}
\end{table}

In most results on a single size network we used $n= 200$ and $k= 100$.  

\subsection{Limit cycles}
Because of the finite phase space of the CA (technically it is
infinite, but the vertex states only fill a finite interval of
$\mathbb Z$ due to the nature of the network laws), network states
will eventually repeat, which leads to a limit cycle because of the
memory-less dynamics. We tested for the appearance of such limit
cycles for different network size $n$ and to our surprise, networks of
type I had with very few exceptions extremely short limit cycles of
period $6$. The exceptions we were able to find, had periods of a
multiple of $6$, the longest found (in a network with $n= 810$) was
$36$. The prevalence of such short limit cycles is still an open
question and beyond this work. We note in this context that already
S. Kauffmann observed such short cycles in his investigation of
switching nets (\cite{Kauffmann1}, \cite{Kauffmann2}) and found it
very amazing. Such short cycles were also found in random networks
(\cite{Huerta}) in a quite different context.

\begin{figure}
  a) \hspace{0.45\textwidth} b) \\
  \parbox[c]{0.55\textwidth}{\small
\begin{tabular}[tb]{>{$}c<{$}|>{$}r<{$}>{$}r<{$}>{$}r<{$}|>{$}r<{$}>{$}r<{$}>{$}r<{$}>{$}r<{$}}
& \multicolumn{3}{c|}{(1)} & \multicolumn{3}{c}{(2)} \\
  t & s_1 & s_2 & J_{12} & s_1 & s_2 & J_{12} \\
\hline
t_0    & 0  & 0  & 1 & 1  & 0  & 1 \\
t_0 +1 & -1 & 1  & 0 & 0  & 1  & 1  \\
t_0 +2 & -1 & 1  & -1 & -1 & 2  & -1 \\
t_0 +3 & 0  & 0  & -1 & 0  & 1  & -1 \\
t_0 +4 & 1  & -1 & 0 & 1  & 0  & -1 \\
t_0 +5 & 1  & -1 & 1 & 2  & -1 & 1 \\
t_0 +6 & 0  & 0  & 1 & 1  & 0  & 1 \\
\hline
\end{tabular}}
  \parbox[c]{0.45\textwidth}{
\includegraphics[width= 0.43\textwidth]{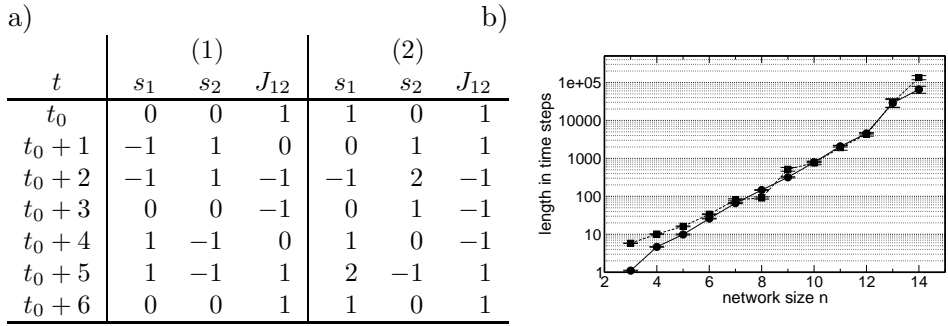}}
\caption{Limit cycle dynamics of a minimal network of $2$
  vertices. \label{table2} b) Length of transients and limit cycles in
  networks of type II.}
\end{figure}

This phenomenon is remarkable in the face of the huge accessible phase
spaces of typical models and points to some hidden ordering tendencies in
these model classes. What is even more startling is that this phenomenon
prevails also in our case for model class $1$ when we introduce a further
element of possible disorder by allowing edges to be dynamically created and
deleted. We formulate the following hypothesis.
\begin{conj}We conjecture that this important phenomenon has its roots
  in the self-referential structure (feed-back mechanisms) of many of the
  used model systems.
\end{conj}
It is instructive to observe the emergence of such short cycles in very small
models on paper, setting for example $\lambda_1=\lambda_2 =0$, i.e., no
switching-off of edges and taking $n=2,3$ or $4$. Taking, e.g., $n=2$ and
starting from $s_2(0)= s_1(0) \mod 2$, the network will eventually reach a
state $s_1(t_0)= s_2(t_0)$. Without loss of generality we can assume
$s_1(t_0) = s_2(t_0)= 0$ and $J_{12}(t_0)=1$.  This state develops into a
cycle of length 6 as illustrated in table \ref{table2}a. For $s_1(0)=
s_2(0)+1 \mod 2$ the state eventually becomes $s_1(t_1) = s_2(t_1) +1$,
without loss of generality $s_1(t_1) = 1$, $s_2(t_1)= 0$, $J_{12}(t_1) = 1$,
resulting in the dynamics in table \ref{table2}b. Again, the length of the
cycle is $6$. Hence, $6$ is a good candidate for a short cycle length, which
-- of course -- does not explain why such a short length should appear at
all.

The transients in networks of type I are also rather short and grow slowly
with the network size (data not shown).

Networks of type II have much longer limit cycles and transients. Because of
numerical limitations we were only able to determine cycle lengths for small
networks. As shown in figure \ref{table2}b) the typical transient and cycle
lengths both grow approximately exponentially.

\subsection{Vertex degrees and internal states}
\begin{figure}
  \includegraphics[width= \textwidth]{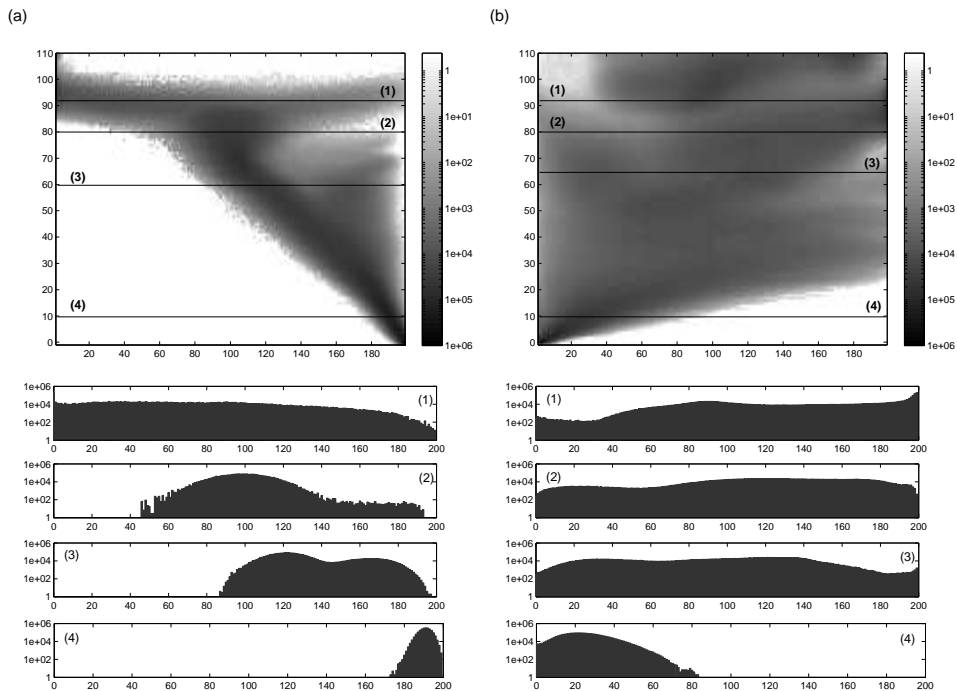}
  \caption{Distribution of abundance of node degrees in dependence on
    $\lambda_1$ (a: network type I, b: network type II). The second
    parameter was fixed to $\lambda_2= 1.2 \cdot \lambda_1$, $n=200$,
    and the total abundances shown are collected over 10990 different
    initial conditions. I.e., for each $\lambda_1$ ($y$ axis) the
    colors of the pixels show the number of times the node degree
    (corresponding to the $x$ position) appeared in the set of graphs
    generated with 10990 initial conditions. Note the logarithmic
    scale on the color scale and the y axes of the examples.
    \label{figurex}}
\end{figure}

To characterize the networks resulting from the two different evolution laws,
we measured several key quantities, including the distribution of node
degrees (also called the \tit{vertex degree sequence}), the distribution of
node states $x_i$, the distribution of bond states $J_{ij}$, as well as
temporal and spatial fluctuations of these quantities. The vertex degree
distribution in dependence on $\lambda_1$ for a network of $200$ vertices is
shown in figure \ref{figurex}a) and b) for network law I and II
respectively. The second parameter was fixed as $\lambda_2= 1.2\cdot
\lambda_1$ to implement a reasonable hysteresis in the dynamical addition and
removal of edges and the degrees were observed after a transient of $10$ time
steps, i.e., prevalently still in a transient dynamics regime. The network
structure undergoes a series of changes for increasing $\lambda_{1/2}$.

Networks of type I evolve from almost fully connected simplex networks to more
sparse connectivities with increasing $\lambda_{1/2}$. There is a regime,
where few vertices with very high degree coexist with many vertices with a
low degree (around $\lambda_1= 60$), which is reminiscent of the situation in
small world networks. We, however, observe a bimodal distribution (with very
sharp peaks in each given network, see figure \ref{figurex}) rather than a
power law of abundance of node degrees. For large $\lambda_{1/2}$ the network
eventually breaks apart and all nodes have vertex degree $0$.

For networks of type II the situation is -- as expected -- inverse with
respect to $\lambda_{1/2}$. The networks are trivial with vertex degree $0$
for all nodes for small $\lambda_{1/2}$ and connect increasingly dense for
increasing $\lambda_{1/2}$. In this family of networks, the distribution of
vertex degrees is always fairly broad and remains such up to large
$\lambda_{1/2}$. We observe an intriguing structure of multiple maxima of the
distributions in a wide range of $\lambda_{1/2}$ values.

\begin{figure}
  \includegraphics[width= \textwidth]{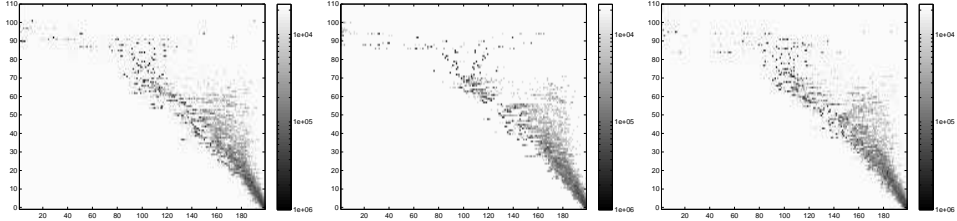}
  \caption{Distribution of abundance of node degrees in network I for
    three separate examples of initial conditions, dependent on
    $\lambda_1$ (again $\lambda_2= 1.2 \cdot \lambda_1$ and $n=200$).
    Note again the logarithmic color scale. All examples reflect the
    overall structure of the averaged data, thus showing that the
    appearance of hubs with large vertex degree and vertices with
    small degree actually occurs within the same network and at the
    same time and is not an artifact of averaging over many initial
    conditions \label{figurey}}
\end{figure}

The observed structure of groups of highly connected and less connected
vertices in the averaged distributions may arise from each given network
realization having these two groups of vertices or appear due to the
existence of two different types of networks. To probe these possibilities, we
examined the abundance distribution of node degrees for individual initial
conditions (Fig. \ref{figurey}). In all examples we observe the same
structure as in the averaged picture (Fig. \ref{figurex}a), such that we have
to conclude that there is only one type of network for a given
$\lambda_1,\lambda_2$ pair that has a structured vertex degree
distribution. Furthermore, careful examination shows that this distribution
can -- unlike in the averaged picture -- be fairly sharp, with often only one
or two prevalent values for the vertex degree. The same is true for networks
of type II: Distributions resemble the averaged picture but often with sharp
peaks for a single value for the vertex degree (data not shown).

The temporal fluctuations, $\text{deg}_i(t+1)-\text{deg}_i(t)$, of vertex
degrees give us insight into the stability of the network structure. For
network law I we observe at $\lambda_1 \approx 35$ an abrupt phase-transition
from basically no temporal fluctuations in the node degrees (``frozen
network'') to fairly high fluctuations (``liquid network''). For even larger
$\lambda$, the fluctuations slowly abate in agreement with the smaller
overall vertex degrees. It is surprising that the transition is so abrupt
especially in the face of the much smoother development of the distribution
of vertex degrees (Fig. \ref{figurex}).

\begin{figure}
  \includegraphics[width= \textwidth]{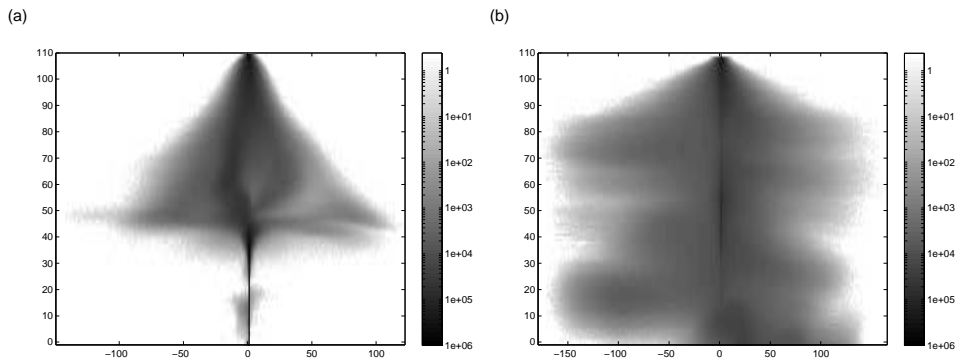}
  \caption{Distribution of abundance of temporal changes in node degrees
  dependent on $\lambda_1$ (again $\lambda_2= 1.2 \cdot \lambda_1$ and
  $n=200$). Note again the logarithmic color scale. \label{figurez}}
\end{figure}
\begin{figure}
  \includegraphics[width= \textwidth]{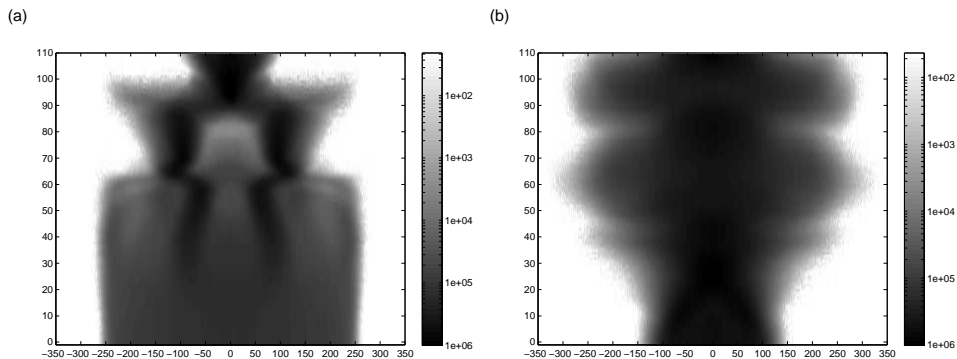}
  \caption{Distribution of abundance of node states in dependence on $\lambda_1$ (again $\lambda_2= 1.2 \cdot \lambda_1$ and
  $n=200$). As before, each line corresponds to a fixed pair of $\lambda_1,
  \lambda_2$ and shows a distribution of observed node states in a
  logarithmic color code. Dark red pixels correspond to node states that are
  observed very often, and dark blue are states that are never observed.
  \label{figurendval}}
\end{figure}

While the node degrees instruct us on the geometry of the CA graph, i.e.,
presumably on the structure of space, the state variables $s_i$ tell us
something about an ``energy'' or ``mass'' density in some appropriate
sense. As explained above the sum over local states, $\sum_I s_i$, is
conserved, and hence is the sum of changes $\sum_I \Delta s_i = 0$.  This
does, however, not imply that the distribution of $s_i$ or changes in $s_i$
are trivial. Figure \ref{figurendval} shows the maps for different node value
distributions depending on $\lambda_1 \in \{1, \ldots, 111\}$ and $\lambda_2=
1.2 \lambda_1$.

The results are, as expected, very different for the two network types.
Networks of type I have a clear region of intermediate $\lambda$ values where
the distribution of node states is strongly bimodal for $62 \leq \lambda_1
\leq 85$. For smaller $\lambda$, i.e., for less bonds being switched off,
there is a broad distribution of node states, and for larger $\lambda$ values,
we observe a sharp unimodal distribution around $0$ corresponding to
disconnected graphs, in which the node states are basically frozen in their
initial values. Intriguingly, the transitions between the different types of
state distributions occur at sharp values of $\lambda$ reminiscent of phase
transitions.

Networks of type II have no bimodal distributions of node values, but there
is a visible modulation in the width of node state distributions with
different values of $\lambda_1$. For some $\lambda$ values, the distribution
of observed states is rather sharp, for others rather wide.  It is also
interesting to note that, contrary to the naive expectation, the total width
of the distributions is slightly larger than for networks of type I (note the
light blue left and right tails for $\lambda_1 \sim 60$). The fact that nodes
with similar node state are connected and can reach some equilibrium and
nodes with very different states are separate and can not interact
(equilibrate) directly seems to allow such ``outlier'' states.

\section{Dimensionality}
When studying such models as described above, it is of tantamount
importance to find certain effective geometric characteristics of the
large scale behavior of such irregular (in the small) networks. This
holds the more so if one plans to perform some coarse-graining or
continuum limit. It is evident that, in particular in the latter
situation, only global features can be significant while, on the other
hand, the finer details on small scales should be ignored.

The notion of \tit{dimension} is one of these fundamental organizing
concepts. There exist of course quite a few different versions of this
concept in the mathematical literature, most of which, while looking
useful at first glance, have to be dismissed on second thoughts. We
will not give the pros and cons of all the different notions (a more
detailed discussion can e.g. be found in \cite{Nowrequ2} and
\cite{Requ2}), let us only make the following point clear. The
discrete models we are dealing with are not to be understood as some
sort of tessellation of a preexisting continuous background manifold
like in algebraic topology. In that case it would for example be
reasonable to use the dimensional concept employed in simplicial
complexes.

In our case, exactly the opposite is true. We view the continuum as an
emergent limit structure, being the result of a (complicated)
renormalisation group like process of coarse-graining. In this
process, due to the rescaling of geometric scales, more and more
vertices and links are absorbed in the \tit{infinitesimal
  neighborhoods} of the emerging points of the \tit{continuum}. As a
consequence, what amounts to a \tit{local} definition of dimension in
the continuum is actually a large scale concept on the network,
involving practically infinitely many vertices and their wiring.

Halpern and Ilachinski make a certain suggestion in their work (see
e.g. \cite{Halpern}), which also happens to be a distinctly local
concept. They define the \tit{effective dimensionality} as the ratio
of the number of next-nearest neighbors to the number of nearest
neighbors averaged over the set of sites of the network. It is a
typical property of most of the different dimensional concepts that
they usually coincide on very regular spaces. This is for example
well-known for the various notions of fractal dimension (cf. e.g.
\cite{Edgar} or \cite {Falconer}). For the notion introduced above we
have for instance in the case of $\Z^2$: $N_{nn}(x)=8,N_n(x)=4$, i.e.
$D_{IH}=2$ and correspondingly for higher dimensions. However, for
other lattices which are not so regular or translation invariant, i.e.
having a more complicated local neighborhood structure, this is no
longer true. While one would still like to associate on physical
grounds in many cases their dimension with the corresponding embedding
dimension of the ambient space, the above quotient may yield a
different value.

The deeper reason is that the accidental near-order of the lattice may
differ from its more important far-order. This phenomenon and the
following physical argument motivated us to choose a different notion
of \tit{intrinsic dimension} which has a lot of very nice and
desirable properties as has been shown in the papers we cited above.
Originally we were primarily motivated by the following reasoning.
What kind of intrinsic global property (i.e. being independent of some
embedding dimension or accidental near-order) is relevant for the
occurrence of phase transitions, critical behavior and the like?  We
wound up with the following answer: It is the increase of number of
new agents or degrees of freedom one sees when one starts from a given
lattice site and moves outward. This led to the following definition (
note that we introduce two slightly different notions which again
coincide in many cases).
 \begin{defi}The (upper,lower) internal scaling dimension with respect
  to the vertex $x$ is given by
\begin{equation}\overline{D}_s(x):=\limsup_{r\to\infty}(\ln\beta(x,r) 
/\ln r)\;,\,\underline{D}_s(x):=\liminf_{r\to\infty}(\ln\beta(x,r)/\ln r)
\end{equation}
The (upper,lower) connectivity dimension is defined correspondingly as
\begin{equation}\overline{D}_c(x):=\limsup_{k\to\infty}(\ln\partial\beta(x,k)
/\ln k)+1\;,\,\underline{D}_k(x):=\liminf_{k\to\infty}(\ln\beta(x,k)/\ln
k)+1
\end{equation}
If upper and lower limit coincide, we call it the internal scaling
dimension, the connectivity dimension, respectively.
\end{defi}
In \cite{Requ2} we exhibited the close connection of this concept with
important properties in various fields of pure mathematics
(\tit{growth properties} of metric spaces). Furthermore, when
performing some continuum limit we could show that our notion of
dimension makes contact with the various notions of \tit{fractal} or
\tit{Haussdorff}-dimension.  
\section{Discussion}
In the preceding sections we introduced and studied networks which are
in various respects generalisations of structural dynamic CA, i.e.,
networks with both the site states and the wiring (link-distribution)
being dynamic and (clock) time dependent. We thus realize an entangled
dynamics of geometry and matter degrees of freedom as in, say, general
relativity or quantum gravity. In contrast to more ordinary CA we
admit a very high degree of disorder in the small while we hope that
our network models find an attracting set in phase space after some
transient time, thus displaying some patterns of global order. We are
particularly interested in ordering phenomena on the geometric side.
That is, we look for collective geometric properties like, e.g.,
dimension, which suggest that, after some coarse graining and/or
rescaling, our networks display global smoothness properties
which may indicate the transition into a continuum-like macro state.

We underpin our investigation by a quite detailed quantitative
computational analysis of various (large scale) characteristics of our
model networks as, e.g., vertex degree distribution, fluctuation
patterns in site and/or link states, etc. There are indications that
for certain choices of the parameters, labelling our model networks,
we witness something akin to structural phase transitions.

It is particularly noteworthy that one of our model networks (after a very
short transient time) enters a periodic state of period only six, and this
being practically independent of the chosen initial state. Given the huge
possible local fluctuations in both site states and link distribution the
extremely short period is remarkable. This phenomenon has also been observed
in other kinds of networks but nevertheless remains somewhat mysterious.


\end{document}